\documentclass[pre,aps,superscriptaddress,floats,showpacs,10pt]{revtex4-1}
\usepackage{graphicx}
\usepackage{dcolumn}
\usepackage{bm}
\usepackage{amssymb}
\usepackage{amsmath}

\begin{document}

\title{Lorentz-Abraham-Dirac vs Landau-Lifshitz radiation friction force in
the ultrarelativistic electron interaction with electromagnetic wave (exact
solutions)}
\author{Sergei V. Bulanov}
\altaffiliation[Also at ]{Prokhorov Institute of General Physics, Russian Academy of Sciences, Moscow
119991, Russia}
\affiliation{Kansai Photon Science Institute, JAEA, Kizugawa, Kyoto 619-0215, Japan}
\author{Timur Zh. Esirkepov}
\affiliation{Kansai Photon Science Institute, JAEA, Kizugawa, Kyoto 619-0215, Japan}
\author{Masaki Kando}
\affiliation{Kansai Photon Science Institute, JAEA, Kizugawa, Kyoto 619-0215, Japan}
\author{James K. Koga}
\affiliation{Kansai Photon Science Institute, JAEA, Kizugawa, Kyoto 619-0215, Japan}
\author{Stepan S. Bulanov}
\altaffiliation[Also at ]{Institute of Theoretical and Experimental Physics, Moscow 117218, Russia}
\affiliation{University of California, Berkeley, CA 94720, USA}

\date{20.10.2011}

\begin{abstract}
When the parameters of electron - extreme power laser interaction enter the
regime of dominated radiation reaction, the electron dynamics changes
qualitatively. The adequate theoretical description of this regime becomes
crucially important with the use of the radiation friction force either in
the Lorentz-Abraham-Dirac form, which possesses unphysical runaway solutions,
or in the Landau-Lifshitz form, which is a perturbation valid for
relatively low electromagnetic wave amplitude. The goal of the present paper
is to find the limits of the Landau-Lifshitz radiation force applicability
in terms of the electromagnetic wave amplitude and frequency. For this, a class of the
exact solutions to the nonlinear problems of charged particle motion in the
time-varying electromagnetic field is used.
\end{abstract}

\maketitle

\section{Introduction}

As with no other problem in classical electrodynamics the problem of the
radiation friction effects on charged particle dynamics has been
attracting attention for more than a century \cite{1, 2, 3, 4}. The
radiation friction imposes constraints on the highest attainable energy of
charged particles accelerated by standard accelerators \cite{5} and in space 
\cite{6}, in particular, on the energy of the ultra high energy cosmic rays 
\cite{7}. The effects of radiation reaction on electrons in a magnetically 
confined plasma lead to the phase space contraction \cite{7a}.
Laser light being coherent and of ultra high intensity
provides special conditions for experimentally studying the radiation
friction effects. The radiation generated by present day \cite{8, 9} lasers
approaches limits when the radiation friction force will change the scenario
of the electromagnetic (EM) wave interaction with matter, i.e. 
at $I>I_{rad}=10^{23}$W/cm$^{2}$. The electron dynamics become dissipative with
fast conversion of the EM wave energy to hard EM radiation, which for
typical laser parameters is in the gamma-ray range \cite{10, 11, 12}. There
are discussions of the modification of the electron acceleration 
in the laser wake field acceleration
regime \cite{12a} and the ion acceleration in the radiation pressure
dominated regime \cite{13} due the radiation friction, which are mainly
obtained with computer simulations \cite{10, 14}. If the laser intensity
substantially exceeds $I_{rad}$, novel physics of abundant electron-positron
pair creation will come into play \cite{15} (see also \cite{12} 
and \cite{15nim}) when the electron (positron) interaction with the EM field is
principally determined by the radiation friction effects. The persistent interest 
towards the radiation friction effects stems from all these reasons
\cite{16, 17}. 

In order to self-consistently find the trajectory of the emitting electron, the so
called Minkovsky equations \cite{3} should be solved with the radiation
friction force taken into account 
\begin{equation}
m_{e}c\frac{du^{\mu }}{ds}=\frac{e}{c}F^{\mu \nu }u_{\nu }+g^{\mu },
\label{eq01a}
\end{equation}%
\begin{equation}
\frac{dx^{\mu }}{ds}=u^{\mu }.  \label{eq01b}
\end{equation}%
Here $u_{\mu }=(\gamma ,\mathbf{p}/m_{e}c)$ is the four-velocity, $F_{\mu
\nu }=\partial _{\mu }A_{\nu }-\partial _{\nu }A_{\mu }$ is the EM field
tensor with $A_{\mu }$ being the EM four-vector and $\mu =0,1,2,3$, and 
\begin{equation}
s=c\int \frac{dt}{\gamma }. 
\end{equation}%
The radiation friction force in the
Lorentz-Abraham-Dirac (LAD) form \cite{A,L,D} is given by 
\begin{equation}
g^{\mu }=\frac{2e^{2}}{3c}\left[ \frac{d^{2}u^{\mu }}{ds^{2}}-u^{\mu }\left( 
\frac{du^{\nu }}{ds}\right) \left( \frac{du_{\nu }}{ds}\right) \right] .
\label{eq02}
\end{equation}%
As is well known, equation (\ref{eq01a}) with the radiation friction
force in the LAD form (\ref{eq02}) possesses unphysical self-accelerating
solutions (e.g. see Refs.  \cite{2, 3}). When the radiation friction force
is taken to be in the Landau-Lifshitz (L-L) form, 
\begin{equation}
g^{\mu }=\frac{2e^{3}}{3m_{e}c^{3}}\left\{ \frac{\partial F^{\mu \nu }}{%
\partial x^{\lambda }}u_{\nu }u_{\lambda }-\frac{e}{m_{e}c^{2}}\left[ F^{\mu
\lambda }F_{\nu \lambda }u^{\nu }-\left( F_{\nu \lambda }u^{\lambda }\right)
\left( F^{\nu \kappa }u_{\kappa }\right) u^{\mu }\right] \right\} ,
\label{eq03}
\end{equation}%
the electron motion equations do not have pathological solutions, 
although they are not always consistent with energy-momentum conservation 
for an abruptly changing electromagnetic field \cite{18a}.

In the 3-dimensional form it can be written as \cite{2} 
\begin{eqnarray}
\mathbf{f} &=&\frac{2e^{3}}{3m_{e}c^{3}\displaystyle{\sqrt{ 1-\frac{v^{2}}{c^{2}}}}}\left\{ \left( \frac{\partial }{\partial t}+
(\mathbf{v} \cdot \bigtriangledown )\right) \mathbf{E}+\frac{1}{c}\left[ \mathbf{v} \times\left( 
\frac{\partial }{\partial t}+(\mathbf{v} \cdot \bigtriangledown )\right) \mathbf{B}%
\right] \right\} +  \notag \\
&&+\frac{2e^{4}}{3m_{e}^{2}c^{4}}\left\{ \mathbf{E\times B}+\frac{1}{c}%
\left( \mathbf{B}\times \left( \mathbf{B\times v}\right) \right) +\frac{1}{c}%
\mathbf{E}\left( \mathbf{v\cdot E}\right) \right\} -  \label{eq03a} \\
&&-\frac{2e^{4}}{3m_{e}^{2}c^{5}\displaystyle{\left( 1-\frac{v^{2}}{c^{2}}\right) }}
\mathbf{v}\left\{ \left( \mathbf{E}+\frac{1}{c}\mathbf{v\times B}\right) ^{2}-\frac{1%
}{c^{2}}\left( \mathbf{v\cdot E}\right) ^{2}\right\} .  \notag
\end{eqnarray}

The L-L radiation friction force being a perturbation is valid provided
there exists a frame of reference, where it is small compared to the Lorentz
force, $eF^{\mu \nu }u_{\nu }$, as noted in Ref. \cite{2}. Proving this
frame of reference existence and finding the range of validity of the friction
force in the L-L form is far from trivial. Below, using several exact
analytical solutions to the electron motion equations in the EM field for
the radiation friction force in the LAD and L-L forms, we discuss the
validity range of the later approximation.

The electron motion equations with the LAD friction force admit an exact
solution for the stationary problem describing the electron motion in the
rotating electric field (see Refs. \cite{18, 11, 12}). This problem can also
be solved for the case of the L-L force. Generalizing the electromagnetic
field configuration, we consider the electric and magnetic field to be a
superposition of components, which are rotating with frequency $\omega $, homogeneous in
space, and time-independent  
\begin{equation}
\mathbf{E}=-\mathbf{e}_{1}E_{1}-\mathbf{e}_{2}
[Dx_{2}+E\cos (\omega t)]-\mathbf{e}_{3}\mathbf{[}Dx_{3}+E\sin (\omega t)],  \label{eq04a}
\end{equation}%
\begin{equation}
\mathbf{B}=\mathbf{e}_{1}B_{1}+\mathbf{e}_{2}Jx_{3}-\mathbf{e}_{3}Jx_{2},
\label{eq04b}
\end{equation}%
where $\mathbf{e}_{1}$, $\mathbf{e}_{2}$ and $\mathbf{e}_{3}$ are the unit
vectors along the 1,2,3 axis. The EM field tensor is equal to%
\begin{equation}
F^{\mu \nu }=\left( 
\begin{array}{cccc}
0 & E_{1} & Dx_{2}+E\cos (\omega t) & Dx_{3}+E\sin (\omega t) \\ 
-E_{1} & 0 & -Jx_{2} & Jx_{3} \\ 
-Dx_{2}-E\cos (\omega t) & Jx_{2} & 0 & -B_{1} \\ 
-Dx_{3}-E\sin (\omega t) & -Jx_{3} & B_{1} & 0%
\end{array}%
\right) .  \label{eq05a}
\end{equation}%
In the case when $E_{1},B_{1},J$ and $D$ vanish, the electric field can be
realized in the antinodes, where the magnetic field vanishes, of a standing
EM wave formed by two counter-propagating circularly polarized EM waves.
Such an EM field configuration plays an important role in theoretical
considerations of various nonlinear effects in quantum electrodynamics, e.g.
see Refs. \cite{12,15,19}. This EM configuration corresponds also to the
circularly polarized EM wave propagating in the underdense plasma for the
frame of reference moving with the wave group velocity \cite{20,21}. In
this frame of reference, the EM wave frequency is equal to the Langmuir
frequency, $\omega _{pe}=\sqrt{4\pi n_{0}e^{2}/m_{e}}$, 
where $n_0$ is the plasma density
and the wave has no
magnetic field component. The static component of the magnetic field, 
$B_{1}$, can be generated in laser plasmas due to the inverse Faraday effect.
Its effect on the charged particle motion has been studied 
in Ref. \cite{18}. The radial component of the electric field, 
$\mathbf{e}_{2}Dx_{2}+\mathbf{e}_{3}Dx_{3}$, and azimuthal component of the magnetic field, 
$\mathbf{e}_{2}Jx_{3}-\mathbf{e}_{3}Jx_{2}$, correspond to a plasma wave in the
boosted frame of reference with $E_{1}$ being the longitudinal component of
the wake field.

It is convenient to write the electron momentum 
$\mathbf{p}=\mathbf{e}_{1}p_{1}(t)+\mathbf{e}_{2}p_{2}(t)+\mathbf{e}_{3}p_{3}(t)$ and coordinates 
$\mathbf{x}=\mathbf{e}_{1}x_{1}(t)+\mathbf{e}_{2}x_{2}(t)+\mathbf{e}_{3}x_{3}(t)$,
as a combination of vectors, which are non-rotating and rotating with angular
frequency $\omega $, 
\begin{equation}
\left( 
\begin{array}{c}
\tilde{u}_{1} \\ 
\tilde{u}_{2} \\ 
\tilde{u}_{3}%
\end{array}%
\right) \equiv \frac{1}{m_{e}c}\left( 
\begin{array}{c}
p_{1} \\ 
p_{||} \\ 
p_{\perp }%
\end{array}%
\right) =\frac{1}{m_{e}c}\left( 
\begin{array}{ccc}
1 & 0 & 0 \\ 
0 & \cos (\omega t) & \sin (\omega t) \\ 
0 & -\sin (\omega t) & \cos (\omega t)%
\end{array}%
\right) \left( 
\begin{array}{c}
p_{1} \\ 
p_{2} \\ 
p_{3}%
\end{array}%
\right)   \label{eq06a}
\end{equation}%
and 
\begin{equation}
\left( 
\begin{array}{c}
\tilde{x}_{1} \\ 
\tilde{x}_{2} \\ 
\tilde{x}_{3}%
\end{array}%
\right) =\left( 
\begin{array}{ccc}
1 & 0 & 0 \\ 
0 & \cos (\omega t) & \sin (\omega t) \\ 
0 & -\sin (\omega t) & \cos (\omega t)%
\end{array}%
\right) \left( 
\begin{array}{c}
x_{1} \\ 
x_{2} \\ 
x_{3}%
\end{array}%
\right) .  \label{eq06b}
\end{equation}

\section{Stationary solution of the electron equations of motion with the
radiation friction force in the Lorentz-Abraham-Dirac form}

The stationary solution of equations (\ref{eq01a}) and (\ref{eq01b}), for which
the vectors $\mathbf{\tilde{u}}=(\tilde{u}_{1},\tilde{u}_{2},\tilde{u}_{3})$ and 
$\mathbf{\tilde{x}}=(\tilde{x}_{1},\tilde{x}_{2},\tilde{x}_{3})$ do not depend on time, with
the radiation friction force in the LAD form, (\ref{eq02}) can be cast as
\begin{equation}
0=a_{1}-\varepsilon _{rad}\tilde{u}_{1}\gamma \left( \gamma ^{2}-1-\left. 
\tilde{u}_{1}\right. ^{2}\right) ,  \label{eq07a}
\end{equation}%
\begin{equation}
\tilde{u}_{2}=\left( d-b-j\frac{\tilde{u}_{1}}{\gamma }\right) \frac{\tilde{u}%
_{2}}{\gamma }+\varepsilon _{rad}\tilde{u}%
_{3}\gamma \left( \gamma ^{2}-\left. \tilde{u}_{1}\right. ^{2}\right) ,
\label{eq07b}
\end{equation}%
\begin{equation}
\tilde{u}_{3}=\left( d+b-j\frac{\tilde{u}_{1}}{\gamma }\right) \frac{\tilde{u}%
_{3}}{\gamma }+a-\varepsilon _{rad}\tilde{u}%
_{2}\gamma \left( \gamma ^{2}-\left. \tilde{u}_{1}\right. ^{2}\right) ,
\label{eq07c}
\end{equation}%
where we use the relationship between $\tilde{x}_{i}$ and $\tilde{u}_{i}$ ,
with $i=1,2,3$, given by Eq. (\ref{eq01b}), which is 
\begin{equation}
\tilde{x}_{1}=s\gamma \tilde{u}_{1}, \quad \tilde{x}_{2}=\frac{\tilde{u}_{3}c}
{\gamma \omega },
\quad
\tilde{x}_{3}=-\frac{\tilde{u}_{2}c}{\gamma \omega }.
\end{equation}%
Here the dimensionless parameter, 
\begin{equation}
\varepsilon _{rad}=\frac{2e^{2}\omega }{3m_{e}c^{3}},  \label{eq08a}
\end{equation}%
characterizes the radiation damping effect, 
$a_{1}=eE_{1}/m_{e}\omega c$, $a=eE/m_{e}\omega c$, $d=eD/m_{e}\omega c$, $j=eJ/m_{e}\omega c $, 
and $b=eB_{1}/m_{e}\omega c$ are normalized longitudinal and
transverse components of the electric and magnetic field, and $\gamma $ is
the electron relativistic Lorentz-factor equal to 
$\sqrt{1+\tilde{u}_{1}^{2}+\tilde{u}_{2}^{2}+\tilde{u}_{3}^{2}}$.
The parameter $\varepsilon _{rad}$ can also be written as $\varepsilon _{rad}=4\pi r_e/3 \lambda $ or 
$\varepsilon _{rad}=2\omega t_e/3$, where $r_e=e^2/m_e c^2$ is the classical 
electron radius, $t_e=r_e/c$, and 
$\lambda =2\pi c/\omega $.

At first we analyze the most simple case with $B_{1}=J=$ $D=0$. The stationary
solution to equations (\ref{eq01a}) and (\ref{eq01b}), for which the vectors 
$\mathbf{\tilde{u}}=(\tilde{u}_{1},\tilde{u}_{2},\tilde{u}_{3})$ and 
$\mathbf{\tilde{x}}=(\tilde{x}_{1},\tilde{x}_{2},\tilde{x}_{3})$ do not depend on time, with the
radiation friction force in the LAD form (\ref{eq02}) can be cast as%
\begin{equation}
0=a_{1}-\varepsilon _{rad}\tilde{u}_{1}\gamma \left( \gamma ^{2}-1-\tilde{u}_{1}^{2}\right) ,  
\label{eq07aa}
\end{equation}%
\begin{equation}
\tilde{u}_{2}=\varepsilon _{rad}\tilde{u}_{3}\gamma \left( \gamma
^{2}-\tilde{u}_{1}^{2}\right) ,  \label{eq07bb}
\end{equation}%
\begin{equation}
\tilde{u}_{3}=a-\varepsilon _{rad}\tilde{u}_{2}\gamma \left( \gamma
^{2}-\tilde{u}_{1}^{2}\right).  \label{eq07cc}
\end{equation}

Multiplying Eq. (\ref{eq07aa}) by $\tilde{u}_{1}$, Eq. (\ref{eq07bb}) by 
$\tilde{u}_{2}$, and Eq. (\ref{eq07cc}) by $\tilde{u}_{3}$, and adding them,
we obtain 
\begin{equation}
a_{1}\tilde{u}_{1}+a\tilde{u}_{2}=\varepsilon _{rad}\gamma ^{3}\left( \gamma
^{2}-1-\tilde{u}_{1}^{2}\right) ,  \label{eq07dd}
\end{equation}%
which is the zero-component of Eq. (\ref{eq01a}).
The left hand side of this equation is proportional to the work produced by the
electric field in the units of time and the right hand side is proportional
to the energy dissipation rate due to the radiation losses.

Multiplying Eq. (\ref{eq07bb}) by $\tilde{u}_{3}$ and Eq. (\ref{eq07cc}) by 
$\tilde{u}_{2}$, and adding them, we obtain 
\begin{equation}
\tilde{u}_{2}^{2}+\tilde{u}_{3}^{2}=a\tilde{u}_{3}.  \label{eq07ee}
\end{equation}

\subsection{Electron in the rotating electric field}

\begin{figure}[tbph]
\includegraphics[width=7cm,height=5cm]{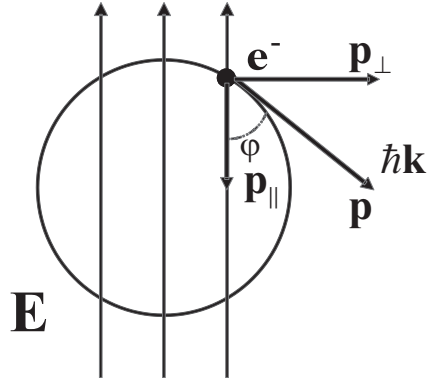}
\caption{ Electron moving in the rotating electric field emits EM radiation.
Due to this the angle between the electron momentum and electric field, 
$\protect\varphi$, is not equal to $\protect\pi/2$.}
\label{FIG01}
\end{figure}

If, in addition, the longitudinal component of electric field vanishes, 
$a_{1}=0$ with $\tilde{u}_{1}=0$, we obtain from Eqs. (\ref{eq07a} - \ref{eq07c}) 
\begin{equation}
p_{||}=\varepsilon _{rad}{p_{\perp }}\gamma ^{3}
\end{equation}
\begin{equation}
p_{\perp }=m_{e}ca-\varepsilon _{rad}{p_{||}}\gamma ^{3},  \label{eq09}
\end{equation}%
where the components of
the electron momentum parallel and perpendicular to the electric field  defined 
by Eq. (\ref{eq06a}) are equal to
\begin{equation}
p_{||}=\frac{(\mathbf{p\cdot E})}{|\mathbf{E}|}=m_{e}c\tilde{u}_{2},
\label{eq09a}
\end{equation}%
\begin{equation} 
 p_{\perp }=\sqrt{p^{2}-p_{||}^{2}}=m_{e}c\tilde{u}_{3},
\label{eq09b}
\end{equation}%
respectively(see Fig. \ref{FIG01}). In this case equation (\ref{eq07ee})
yields a relationship between $p_{||}$ and $p_{\perp }$:

\begin{equation}
p_{||}^{2}+p_{\perp }^{2}=m_{e}cap_{\perp }. \label{eq010}
\end{equation}

The electron gamma-factor $\gamma $ is equal to 
$\sqrt{1+\tilde{u}_{2}^{2}+\tilde{u}_{3}^{2}}\equiv \sqrt{1+
p_{||}^{2}/m_e^2 c^2+p_{\perp }^{2}/m_e^2 c^2}$. 
As we see, from the relationship 
\begin{equation}
p_{||}=\sqrt{p_{\perp }(m_{e}ca-p_{\perp})}
\end{equation}
it follows that the component of the electron momentum perpendicular to the electric field
 is always equal or less than $a$.
Multiplying Eq. (\ref{eq09a}) by $p_{\perp }$ and the
Eq. (\ref{eq09b}) by $p_{||}$ and subtracting them, we find 
\begin{equation}
a\frac{p_{||}}{m_e c}=\varepsilon _{rad}\gamma ^{3} (\gamma^2 -1),  \label{eq011}
\end{equation}%
which corresponds to the energy balance equation (\ref{eq07dd}) for 
$a_{1}=\tilde{u}_{1}=0$ in the limit $a \gg 1$.

If the EM field amplitude is relatively small, i.e. 
$1\ll a\ll \varepsilon_{rad}^{-1/3}$ Eqs. (\ref{eq010}) and (\ref{eq011}) yield for the components 
of the electron momentum perpendicular and parallel to the electric field 
\begin{equation}
p_{\perp }\approx m_{e}c\left( a-\varepsilon _{rad}^{2}a^{7}\right) 
\end{equation}%
\begin{equation}
p_{||}\approx m_{e}c\varepsilon _{rad}a^{4}.  \label{eq012}
\end{equation}%
In the opposite limit, when $a\gg \varepsilon _{rad}^{-1/3}$, we obtain 
\begin{equation}
p_{\perp }\approx \frac{m_{e}c}{\sqrt{\varepsilon _{rad}a}}
\end{equation}%
\begin{equation}
 p_{||}\approx m_{e}c\left( \frac{a}{\varepsilon _{rad}}\right) ^{1/4}.
\label{eq013}
\end{equation}%

\begin{figure}[tbph]
\includegraphics[width=15cm,height=4.5cm]{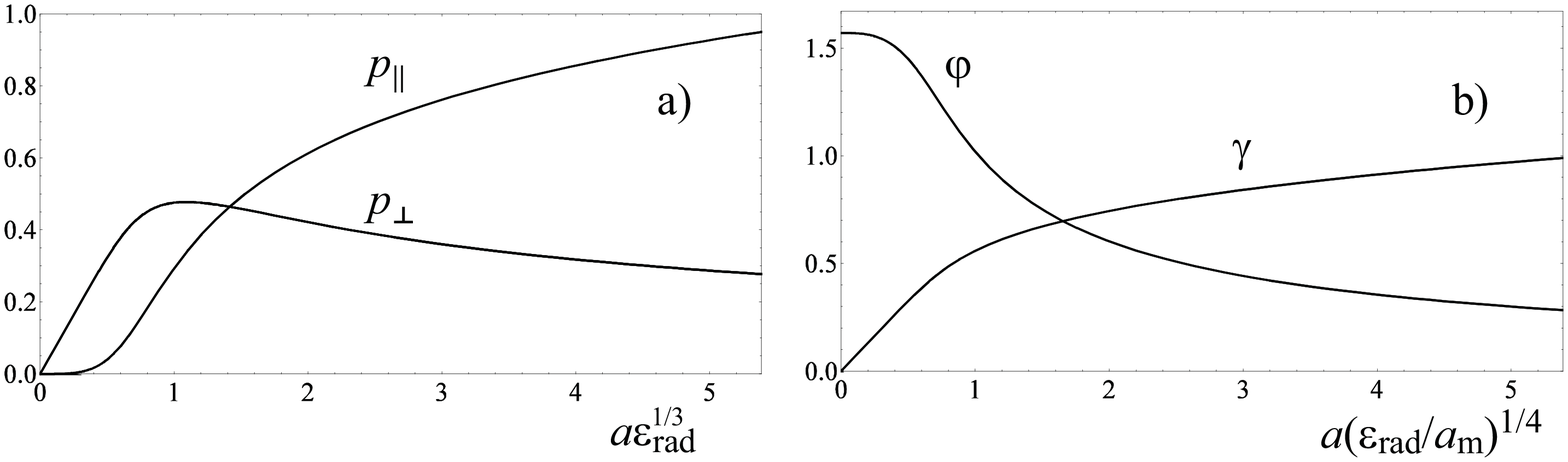}
\caption{ a) Dependence of the components 
of the electron momentum perpendicular, $p_{\perp }$, and parallel, $p_{||}$, 
to the electric field  (normalized by $m_e c (a_m/\varepsilon_{rad})^{1/4}$) 
on the normalized EM field amplitude, 
$a \varepsilon_{rad}^{1/3}$, and b) dependence of $\varphi$ and electron gamma-factor $\gamma$,
normalized by $(a_m/\varepsilon_{rad})^{1/4}$ , 
 on  
$a \varepsilon_{rad}^{1/3}$ for $a_m=2500$ and $\varepsilon_{rad}=10^{-8}$.}
\label{FIG02}
\end{figure}

In Fig. \ref{FIG02}a we show the dependence of $p_{\perp }$ and $p_{||}$ on the
EM field amplitude, $a$, for the dimensionless parameter $\varepsilon _{rad}=10^{-8}$, obtained
by a numerical solution of Eqs. (\ref{eq09}). Here, the horizontal axis is 
normalized by $\varepsilon_{rad}^{-1/3}$
and the vertical axis is normalized by $(a_m/\varepsilon_{rad})^{1/4}$.

As we see, the dependences of the components of 
the electron momentum perpendicular and parallel to the electric
field correspond the asymptotics given by
Eqs. (\ref{eq012}) and (\ref{eq013}). The perpendicular momentum reaches the
maximum at $a\approx \varepsilon _{rad}^{-1/3}$ and then decreases. The
parallel momentum component monotonically increases with the EM amplitude
growth.

It is also convenient to represent the momentum components in the complex
form 
\begin{equation}
p_{||}+ip_{\perp }=\mathfrak{p}\exp (i\varphi )  \label{eq014}
\end{equation}%
with $\mathfrak{p}=\sqrt{p_{\perp }^{2}+p_{||}^{2}}$ and $\varphi $ being
the momentum value and the phase between the rotating electric field and the
momentum vector. Eqs. (\ref{eq09}) can be rewritten as 
\begin{equation}
a^2=\left(\gamma^2-1\right)(1+\varepsilon _{rad}^{2}\gamma ^{6})
\end{equation}%
\begin{equation}
\tan \varphi =\frac{1}{\varepsilon _{rad}\gamma ^{3}},
\label{eq015}
\end{equation}%
where the electron gamma-factor $\gamma $ is equal to 
$\sqrt{1+\mathfrak{p}^{2}}$. These equations are the same as Eqs. (6) in Ref. \cite{12}.

In Fig. \ref{FIG02}b we present the electron gamma factor $\gamma $ normalized
by $(a_{m}/\varepsilon _{rad})^{1/4}$ and the angle $\varphi $ versus the EM
field amplitude $a$ for $\varepsilon _{rad}=10^{-8}$. The angle $\varphi $
changes from $\pi /2$ at $a=0$, when the electron momentum is perpendicular
to the electric field vector, to 0 at $a\rightarrow \infty $, when the
electron momentum becomes antiparallel to the electric field. The horizontal axis is 
normalized in the same way as in Fig. \ref{FIG02}a.

\subsection{Electron in the superposition of rotating and radial electric
fields}

Electron dynamics in the superposition of rotating and radial electric field
corresponds to the case of the electron direct acceleration by the laser
pulse propagating inside the self-focusing channel \cite{27}. Its realization 
provides the conditions for substantial enhancement of the betatron radiation
allowing for photon emission in the gamma ray energy range \cite{27a}. 

In the frame
of reference moving with the laser pulse group velocity, $v_g$, the equations of the 
electron motion are Eqs. (\ref{eq07a} - \ref{eq07c}) with $b=j=a_{1}=0$:%
\begin{equation}
(\gamma-d)\tilde{u}_{2}=\varepsilon _{rad}\tilde{u}_{3}\gamma ^{4},  \label{eq017aa}
\end{equation}%
\begin{equation}
(\gamma-d)\tilde{u}_{3}=a \gamma -\varepsilon _{rad}\tilde{u}_{2}\gamma ^{4}.  \label{eq017bb}
\end{equation}%
For variables $\mathfrak{p}$ and $\varphi $ defined by Eq. (\ref{eq014}) we
can rewrite Eqs. (\ref{eq017aa} , \ref{eq017bb}) as%
\begin{equation}
a^2=
\left(\gamma^{2}-1 \right)
\left[
\left( 1-\frac{d}{\gamma} \right)^{2}
+\varepsilon _{rad}^{2}
\gamma ^{6} 
\right],
\end{equation}%
or
\begin{equation}
d=\gamma-\gamma \sqrt{\frac{a^2}{\gamma^2-1}-\varepsilon _{rad}^{2}\gamma^6}
\label{eq:d}
\end{equation}%
and
\begin{equation}
\tan \varphi =\frac{\gamma -d}{\varepsilon _{rad}\gamma^{4}}
\end{equation}%
with $\gamma =\sqrt{1+\mathfrak{p}^{2}}$. 

\begin{figure}[tbph]
\includegraphics[width=15cm,height=4.5cm]{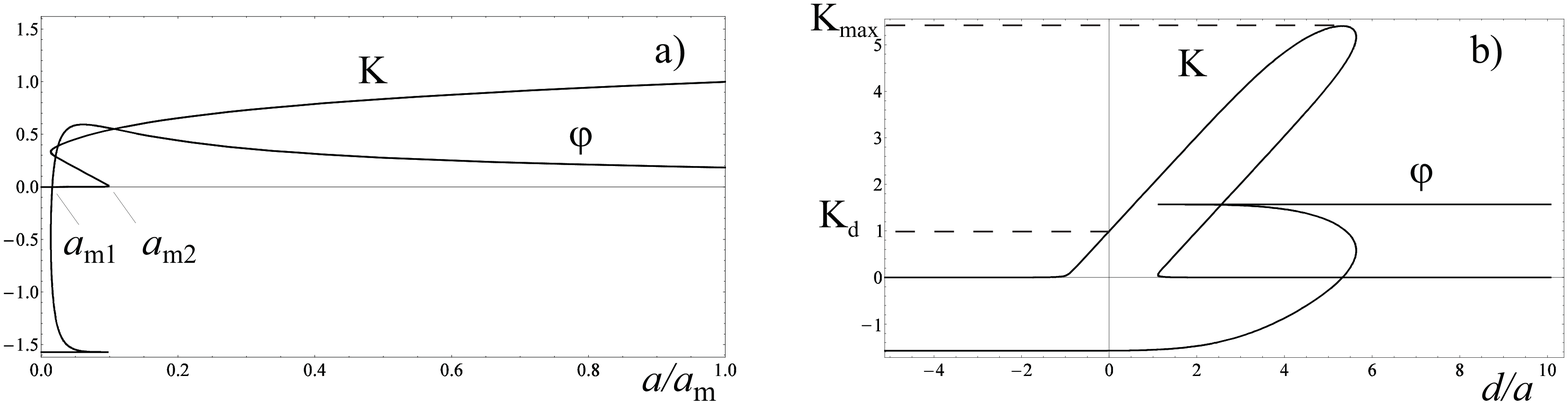}
\caption{ Solution of the electron equation of motion  with the radiation
friction force in the LAD form in the case of the superposition of rotating and radial electric
fields. a) Normalized electron energy,  $K=(\gamma-1)/(\gamma_m -1) $, 
 and the angle, 
$\varphi $, vs the electric field amplitude $a/a_m$ with $a_m=2500$. b) Normalized electron energy,  
$K=(\gamma-1)/(\gamma_m -1) $, 
and the angle, $\varphi $, vs parameter $d/a$ for $a=50$.
Here $\gamma_m=\sqrt{1+(p_m/m_ec)^2}$ with $p_m/m_ec=700$, $d=250$, and $\varepsilon_{rad}=10^{-8}$.}
\label{FIG03}
\end{figure}

In Fig. \ref{FIG03}a we show
dependences of the normalized electron energy,  $K=(\gamma-1)/(\gamma_m -1)$, 
where $\gamma_m=\sqrt{1+(p_m/m_ec)^2}$ with $p_m/m_ec=700$, $d=250$, 
$\varepsilon_{rad}=10^{-8}$, and angle, 
$ \varphi $, on the electric field amplitude $a$. As we see, asymptotically at
$\gamma \rightarrow \infty $ their behaviour is the same as in the above
discussed case corresponding to 
Eqs. (\ref{eq012}, \ref{eq013} and \ref{eq015}) and illustrated by Fig. \ref{FIG02}. 
In the relatively low energy
region the dependence of electron momentum on the electric field amplitude
shows the hysteresis behaviour as seen in  Fig. \ref{FIG03}a. In the
region $a_{m1}<a<a_{m2}$, with 
\begin{equation}
a_{m1}\approx \varepsilon _{rad}d^{4}
\end{equation}%
\begin{equation}
 a_{m2}\approx d,
\label{am1am2}
\end{equation}%
there are three values of the electron momentum corresponding to one value
of $a$. At $a\approx a_{m1}$ the electron energy is approximately equal to 
$\gamma _{m1}\approx d$ and for $a\approx a_{m2}$ we have $\gamma _{m2}\approx d^{1/3}$, 
provided $4 \varepsilon _{rad}d^{3}\ll 1$. 
The condition for the hysteresis to occur is
\begin{equation}
\varepsilon _{rad}d^{2/3}\leq 0.276.
\end{equation}%
The  hysteresis 
is distinctly seen in Fig. \ref{FIG03}b showing typical behaviour for nonlinear resonance \cite{28}
the nonlinear resonance dependence of the quiver energy, $K$, on the parameter $d$,
which is equal to the square of the ratio of the electron oscillation frequency in the radial
electric field to the frequency of the EM wave.
This corresponds to the nonlinear regime of the "betatron resonance" studied
in Refs. \cite{27}. 

For the electron moving inside the self-focusing channel
a typical value of the parameter $d$ is of the order of 
\begin{equation}
d=\frac{\omega _{pe}^2}{\omega^{2}}\gamma _{g},
\end{equation}%
where $\gamma_g=(1-v_g^2/c^2)^{-1/2}$, i.e. it is approximately equal to $\gamma _{g}$. 
Here we take into account that in the boosted frame of reference the EM wave frequency 
is equal to the Langmuir frequency.
As a result we can find the electron energy
accelerated by the "betatron resonance" mechanism in the laboratory frame of
reference to be equal to%
\begin{equation}
\mathcal{E}_{e}\approx m_{e}c^{2}\gamma _{g}^{2}.
\end{equation}%
This corresponds to $K_d$ in Fig. \ref{FIG03}b.
For example, for the plasma density of the order of $10^{19}cm^{-3}$ and the
laser wavelength $\approx 1\mu m$, the electron energy is about 50 MeV. The
transverse component of the electron momentum is of the order of 
$m_{e}c\gamma _{g}\approx 5$MeV.

The maximal energy, which an electron can achieve in the resonance case according to Eq. (\ref{eq:d})
in the boosted frame of reference is approximately equal to 
$\tilde{\cal E}_e=mec^2\gamma_{max}=mec^2(a/\varepsilon_{rad})^{1/4}$, which corresponds to $K_{\rm max}$
in Fig. \ref{FIG03}b. In the laboratory frame of reference 
we obtain 
\begin{equation}
{\cal E}_e\approx mec^2\gamma_{max}\gamma_g.
\end{equation}
For $a\approx 10$ and $\gamma_g=10$ it gives 
${\cal E}_e\approx$ 500 MeV. The  energy of emitted photons is of the order of 
$ \approx 0.3 \hbar \omega \gamma_{max}^3$. This is of the order of 300 KeV.
The emitted $\gamma$-rays are collimated within the angle $\approx \gamma_g$.

\section{Stationary solution of the electron equations of motion with the
radiation friction force in the Landau-Lifshitz form}

Here we analyze the electron motion equations with the radiation friction force taken in
the L-L form (\ref{eq03}) for a stationary electron orbit. We retain the
leading order terms in the limit $\gamma \gg 1$, which is the last term in Eq. (\ref{eq03a}).

\subsection{Electron motion in the rotating homogeneous electric field}

\begin{figure}[tbph]
\includegraphics[width=15cm,height=4.5cm]{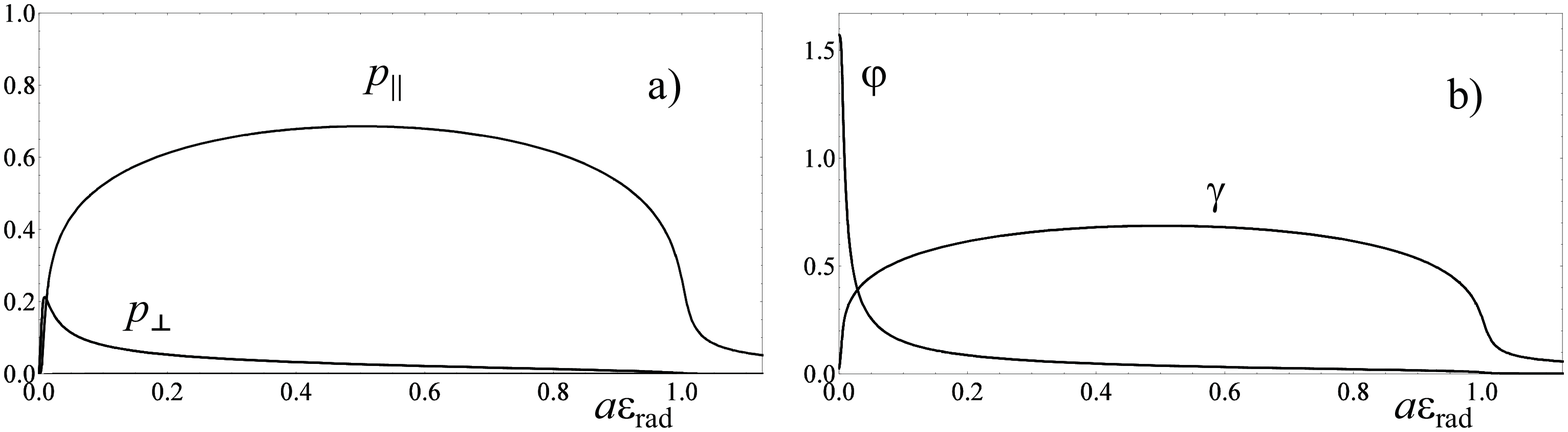}
\caption{ Solution of the electron equation of motion  with the radiation
friction force in the LL form in the case of rotating homogeneous electric field: 
a) Dependence of the components 
of the electron momentum (normalized by $m_e c (a_m/\varepsilon_{rad})^{1/4}$) 
perpendicular, $p_{\perp }$, and parallel, $p_{||}$, to the electric field 
on the normalized EM field amplitude, 
$a \varepsilon_{rad}$, and b) dependence of $\varphi$  and  
the electron gamma-factor, 
$\gamma$ (devided by $(a_m/\varepsilon_{rad})^{1/4}$), on  
$a \varepsilon_{rad}$ for $a_m=1500$ and $\varepsilon_{rad}=7.5 \times 10^{-4}$.}
\label{FIG04}
\end{figure}

We look for the solutions describing a stationary electron orbit in a rotating
homogeneous electric field, i.e. $E_{1},D,B_{1},J$ vanish in 
Eqs. (\ref{eq04a} - \ref{eq05a}). From Eq. (\ref{eq03a}) we obtain for the $\tilde{u}_{2}=p_{||}/m_e c$
and $\tilde{u}_{3}=p_{\perp }/m_e c$ momentum components the algebraic equations 
\begin{equation}
\tilde{u}_{2}=\varepsilon _{rad}\frac{\tilde{u}_{3}}{\gamma }a^{2}\left(1+\tilde{u}_{3}^2\right),
\end{equation}%
\begin{equation}
\tilde{u}_{3}=a-\varepsilon _{rad}\frac{\tilde{u}_{2}}{\gamma }a^{2}\left(1+\tilde{u}_{3}^2\right).  
\label{eq018}
\end{equation}%
Using the variables $\mathfrak{p}$ and $\varphi $ defined by Eq. (\ref{eq014}) 
we can present these equations in the form 
\begin{equation}
a^2=\frac{\gamma^2}{2 \varepsilon _{rad}^2 (\gamma^2-1)}
\left[1-\sqrt{1-4 \varepsilon _{rad}^2 (\gamma^2-1)^2} \right]-(\gamma^2-1)^2,
\end{equation}%
\begin{equation}
 \tan \varphi =\frac{2 \varepsilon _{rad}(\gamma^2-1)}{
\gamma-\sqrt{\gamma^2-4 \varepsilon _{rad}^2 \gamma^2(\gamma^2-1)^2}}.
\label{eq019}
\end{equation}%
In the range of the EM field amplitude, $1 \ll a \ll \varepsilon _{rad}^{-1}$, 
solution to these equations has the same asymptotic dependences as given by
Eqs. (\ref{eq09}) and (\ref{eq010}). However, when the EM field amplitude
approaches the value of $\varepsilon _{rad}^{-1}$, the solution
qualitatively changes. According to Eq. (\ref{eq019}), the electron momentum
decreases as also shown in Fig. \ref{FIG04}. In Fig. \ref{FIG04}a we present
the components of the electron momentum parallel and perpendicular to the instantaneous electric field 
as functions of the electric field amplitude. Fig. \ref{FIG04}b shows the dependences of the 
angle, $\varphi$, and the electron gamma-factor, $\gamma$, on the electric field. 
The momentum and gamma-factor
are normalized by $(a_m/\varepsilon_{rad})^{1/4}$, 
and the dimensionless electric field amplitude by $\varepsilon_{rad}$.

\begin{figure}[tbph]
\includegraphics[width=7.5cm,height=4.5cm]{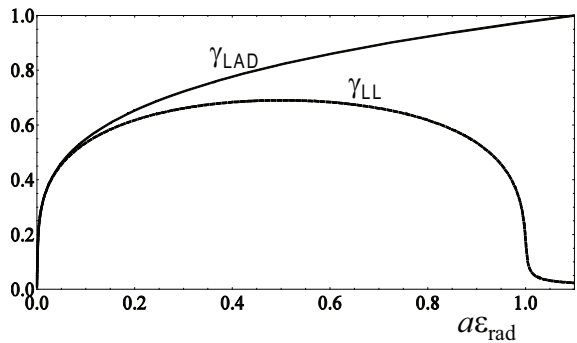}
\caption{ Solution of the electron equation of motion  in a rotating homogeneous electric field
for $a_m=1500$ and $\varepsilon_{rad}=7.5 \times 10^{-4}$.
Dependences of the electron gamma-factors on the electric field: $\gamma_{\rm LAD}$ and $\gamma_{\rm LL}$ 
correspond to the radiation friction force taken in the LAD and L-L form, respectively. 
The normalization is the same as in Fig. \ref{FIG04}.}
\label{FIG05}
\end{figure}

In Fig. \ref{FIG05} we present the results of the solution of the electron motion equation 
in a rotating homogeneous electric field.
Here the dependences of the electron gamma-factors on the electric field, $\gamma_{\rm LAD}$ 
and $\gamma_{\rm LL}$,  correspond to the radiation friction force taken in the LAD and 
L-L form, respectively. 
The normalization is the same as in Fig. \ref{FIG04}. 

\subsection{Electron in the superposition of rotating and radial electric
fields}

The equations of the electron motion in the superposition of rotating and radial electric
fields ($b=j=a_{1}=0$) with the radiation friction force in the LL form
are:
\begin{equation}
\left(\gamma-d \right) \tilde{u}_{2}
=\varepsilon _{rad}\tilde{u}_{3}
\left[\left(d \gamma+ a \tilde{u}_{3}\right)^2+a^2 -d^2 \right],  \label{eq019a}
\end{equation}%
\begin{equation}
\left(\gamma-d\right) \tilde{u}_{3}=a \gamma-\varepsilon _{rad}\tilde{u}_{2}
\left[\left(d \gamma+ a  \tilde{u}_{3}\right)^2+a^2 -d^2 \right].  \label{eq019b}
\end{equation}%
We can rewrite Eqs. (\ref{eq019a} , \ref{eq019b}) as%
\begin{equation}
a^2 =d^2 - \frac{(d - \gamma + \gamma^3)^2}{\gamma^2} +
\frac{\gamma^2-\sqrt{\gamma^4(1-4 \varepsilon _{rad}^2 (\gamma^2-1)^2)}}
{2 \varepsilon _{rad}^2  (1 - \gamma^2)}
\end{equation}%
\begin{equation}
\tan \varphi =\frac{2\varepsilon _{rad}(\gamma -d)(\gamma^2-1)}{\gamma^2 \left(1-\sqrt{1-4 \varepsilon _{rad}^2 (\gamma^2-1)^2} \right)},
\end{equation}%
where  $\varphi $ is defined by Eq. (\ref{eq014}). 

As in the above considered case described by the LAD model, the dependence of electron momentum 
on the electric field amplitude shows similar behaviour, provided $a \gg 1$, $\gamma \gg 1$, and 
$2 \varepsilon _{rad} \gamma^2 \ll 1$.

\section{Discussions}

High order derivatives with respect to time in the electron motion equations with the radiation 
friction force in the LAD form stem from the 3D geometry of the electromagnetic field interaction with 
a point charge, when the electrostatic energy diverges for the charge radius tending to zero leading to 
the so-called classical mass renormalization \cite{3,4}. In a  1D electrodynamics model with 
the point charge  role played by an infinitely thin foil there are no such difficulties 
(see analysis of this case in Appendix I). 

As follows from consideration of the above presented exact solutions to the electron equations 
of motion with the 
radiation friction force taken in the LAD and L-L form, in the limit of relatively low electric
field amplitude they show the same behaviour, as seen in Fig. \ref{FIG05}. 
When the electric field is strong, i.e. the normalised field
amplitude, $a$, approaches the value of $\varepsilon _{rad}^{-1}$, the solutions are drastically different.

The condition $a=\varepsilon _{rad}^{-1}$ corresponds to the electric field
equal to the critical electric field of classical electrodynamics, 
\begin{equation}
E_{cr}=\frac{m_{e}^{2}c^{4}}{e^{3}}.  \label{eq020}
\end{equation}%
This electric field can produce a work equal to $m_{e}c^{2}$ over the distance of the classical 
electron radius $r_{e}$. 

The radiation friction force in the Landau-Lifshitz form assumes the smallness of 
the EM field amplitude compared to the  critical field of classical electrodynamics. 
Another parameter which should be small is the ratio of the EM field inhomogeneity scale length 
to the classical electron radius, $r_e$. 
The time dependent EM fields should be slowly evolved on a timescale compared to $t_e=r_e/c$, 
as discussed in Refs. \cite{18a, 29} devoted to the problem of classical electrodynamics applicability. 

Obviously, the limit of the EM field amplitude of the order of $E_{cr}$ 
(and of the space- and time scales of the order of 
$r_e$ and $t_e=r_e/c$) is of pure academic interest,
because quantum mechanical effects become important at electric field amplitudes substantially 
lower (and on spatial scales of the order of the electron Compton wavelength, $\hbar/m_e c$). 
The critical electric field of quantum electrodynamics (QED),
 \begin{equation}
E_{S}=\frac{m_{e}^{2}c^{3}}{e \hbar},  
\end{equation}
 is a factor
$\alpha=e^2/\hbar c \approx 1/137$ smaller. Here $\alpha$ is the fine structure constant. 
For the electron motion in colliding EM 
waves, the QED effects due to the recoil from the photon emission, 
should be incorporated into the description of the electron 
interaction with the EM field for an even smaller EM wave amplitude. 
As shown in Ref. \cite{11}, the recoil from the photon emission comes into play when the 
photon momentum, $\hbar {\bf k}_m$ becomes of the order of the electron momentum, $ {\bf p}$.
In other words, the photon with the energy larger than the electron cannot be emitted.
Here the wave vector of the photon emitted by the electron, ${\bf k}_m$, 
is of the order of $\omega \gamma^3/c $
and $p\approx m_e c\gamma$. This yields the gamma factor, 
at which the photon recoil should be taken into account, 
equal to $\gamma_{QM}=\sqrt{m_ec^2/\hbar \omega}$. 
As noted above, for $a\gg \varepsilon _{rad}^{-1/3}$ 
there is a relationship between the electron energy and the EM field amplitude, which has a form 
$\gamma=(a/\varepsilon _{rad})^{1/4}$. For $(a/\varepsilon _{rad})^{1/4}=\gamma_{QM}$ we have 
$a=a_{QM}=\varepsilon _{rad} m_e^2 c^4/\hbar^2 \omega^2=2e^2 m_e c/\hbar^2 \omega$.
The electric field, $m_e c \omega a_{QM}/e$, is of the order of $\alpha E_S$, which is 
equal to $\alpha^2 E_{cr}$, below which both the LAD and L-L forms 
for the radiation friction force give the same result.
This confirms a conclusion made in Ref. \cite{29a}, where 
on the basis of numerical calculations of an electron motion in a very
strong laser pulse it was found that in the classical regime the L-L damping equation is very adequate.

Although conclusions following from the above presented consideration do not have the character 
of a rigorously proved mathematical theorem,
they give an indication of the range of the L-L forms for the radiation friction force, 
which can be written in terms of the normalized EM wave 
amplitude:
\begin{equation}
a<a_{QM}.  \label{eq021}
\end{equation}%

We note here that the question as to whether or not the EM field amplitude is small compared to $E_{cr}$ 
should be answered based on the examination of the field in the electron rest frame of reference. 
For example, if a relativistic electron bunch interacts with the 
EM field, the radiation friction force in the L-L form can predict electron behaviour 
different from that described  with the LAD radiation friction
force at substantially lower electric field amplitude.

Quantitatively this limit is described in terms of the Lorentz and gauge invariant parameter
\begin{equation}
\chi=\frac{\sqrt{\left( F^{\mu \nu }u_{\nu}\right)^2}}{E_{cr} }.  \label{eq0chi}
\end{equation}%
It is of the order of the ratio $E=E_{cr}$ in the electron rest frame
of reference. It can be expressed via the electric and magnetic fields and electron momentum as
\begin{equation}
\chi=\frac{
\sqrt{
\left( m_e c \gamma {\bf E}+{\bf p} \times {\bf B} \right)^2 -\left({\bf p}\cdot {\bf E}\right)^2
}
}{
m_e c \, E_{cr} 
}.  
\label{eq00chi}
\end{equation}%
For the case of an electron interacting with a laser pulse, using the solution presented 
in Ref. \cite{2}, we find that the
components of the electron momentum along, $p_1=({\bf p}\cdot {\bf k})/|{\bf k}|$, and perpendicular, 
${\bf p_a}={\bf p} -{\bf k}p_1/|{\bf k}|$, to the direction of the laser pulse propagation can
be found from the equations
\begin{equation}
{\bf p_a}={\bf p}_{{\bf a},0}+m_e c ({\bf a}-{\bf a}_0)
\label{eq0papk1}
\end{equation}%
 and
\begin{equation}
\sqrt{ m_e^2 c^2+[{\bf p}_{{\bf a},0}+m_e c({\bf a}-{\bf a}_0)]^2+p_1^2 } -p_1=
\sqrt{ m_e^2 c^2+[{\bf p}_{{\bf a},0}-m_e c{\bf a}_0]^2+p_{1,0}^2 } -p_{1,0}.  
\label{eq0papk}
\end{equation}%
Here ${\bf k}= |{\bf k}|{\bf e}_1$ is the wave vector of the EM wave.
These expressions use the conservation of generalised momentum, assuming the radiation
friction effects are negligibly small (see analysis of the radiation damping effects 
on the parameter $\chi$ in Appendix II). 
$p_{1,0}$ and ${\bf p}_{{\bf a},0}$ are the initial components of the electron momentum, i.e. before
collision with the laser pulse for ${\bf a}_0=0$. 

For a plane EM wave
propagating along the x-axis with the electric, ${\bf E}=-c^{-1}\partial_t {\bf A}$ 
and magnetic field ${\bf B}=\nabla \times {\bf A}$, where
 ${\bf A}(x - ct)$ is the EM vector potential, prime denotes differentiation 
 with respect to the variable $x - ct$, and ${\bf a}=e{\bf A}/m_e c^2$, the invariant, 
 $\chi$, takes the form
\begin{equation}
\chi=\frac{E}{E_{cr}}
\left( \gamma  -\frac{p_1}{m_e c}\right).  
\label{eq0chi-w}
\end{equation}%
Substituting expression (\ref{eq0papk}) to Eq. (\ref{eq0chi-w}) we obtain
\begin{equation}
\chi=\frac{E}{E_{cr}}
\frac{ \sqrt{m_e^2 c^2+({\bf p}_{{\bf a},0}-m_e c{\bf a}_0)^2+p_{1,0}^2 } -p_{1,0}}{m_e c}.  
\label{eq0chi-wa}
\end{equation}%

As we see, for an ultrarelativistic electron, $p_{1,0}\gg m_e c$,  colliding with 
the laser pulse in the co-propagating configuration, i.e. $p_{1,0}>0$, 
${\bf p}_{{\bf a},0}=0$ and ${\bf a}_0=0$,
the parameter $\chi$ is negligibly small: 
\begin{equation}
\chi \approx (E/E_{cr})(m_e c/2 p_{1,0}).
\end{equation}%
In the case of the head-on collision of an ultrarelativistic electron 
with the laser pulse, when $p_{1,0}<0$ , 
${\bf p}_{{\bf a},0}=0$ and ${\bf a}_0=0$, the parameter $\chi$ is a factor $(2 p_{1,0}/ m_e c)^2$ 
larger and 
is approximately equal to 
\begin{equation}
\chi \approx (E/E_{cr})(|2 p_{1,0}|/m_e c).
\end{equation}%
 If the electron appears inside the laser pulse as result of a gas ionization
or due to the electron-positron pair creation (see Refs. \cite{12,15,29b}), 
the initial electron momentum is negligibly small, $p_{1,0}\approx 0$ and  
${\bf p}_{{\bf a},0} \approx 0$, the constant ${\bf a}_0$ corresponds 
to the EM field in the point and instant of time where and when the electron 
is created. In this case the invariant $\chi$ is equal to 
\begin{equation}
\chi =\frac{E}{E_{cr}}\sqrt{1+{{\bf a}_0}^2}\approx \frac{a a_0}{a_{cr}}
\end{equation}%
 with $a_{cr}=eE_{cr}/m_e \omega_0 c=
1/(\omega_0 t_e)$.

The above defined parameter $\chi$ is again a factor $1/\alpha \approx 137$ smaller 
than the known quantum electrodynamics 
parameter $\chi_e=\sqrt{\left( F^{\mu \nu }u_{\nu}\right)^2}/E_{S} $, which
gives the ratio of the EM field amplitude to the QED critical field, $E_S$, 
in the electron rest frame of reference \cite{30}.
For a 1$\mu$m ten-petawatt laser pulse focused to a few microns focus spot 
with the dimensionless amplitude $a=3\times 10^2$ the
parameter $\chi$ becomes equal to unity for the electron energy of about 
of 40 GeV and  the QED parameter $\chi_e$ 
is of the order of unity
for the electron energy of about 300 MeV.

\medskip
\begin{acknowledgments}
We thank G. Korn, A. Macchi, T. Nakamura, N. B. Narozhny, F. Pegoraro, A. S. Pirozhkov, 
H. Ruhl, M. Tamburini, and A. G. Zhidkov 
for discussions. We acknowledge support of this work from the Ministry of
Education, Culture, Sports, Science and Technology (MEXT) of Japan,
Grant-in-Aid for Scientific Research, No. 20244065.
\end{acknowledgments}

\appendix
\section{ RADIATION FRICTION IN 1D ELECTRODYNAMICS}

In a  1D electrodynamics model with 
the point charge  role played by an infinitely thin foil \cite{31} there 
are no difficulties with high order derivatives 
with respect to time in the electron motion equations with the radiation 
friction force. This model has been extensively used in studying the problem
 of relativistic thin plasma layer transparency, 
 particularly for the purposes of the laser pulse shaping \cite{32} (see also the experimental 
 paper \cite{33}), 
in the high order harmonics generation \cite{34}, in the laser ion acceleration 
\cite{13, 35}, and in the generation of coherent extremely high intensity x-ray pulses 
by relativistic mirrors \cite{36}.

Using the results of Refs. \cite{31, 32}, we consider the case of normal incidence
of a plane electromagnetic wave on an infinitely thin foil. The foil is
 located in the plane $x=0$. The interaction of
the wave with the foil is described by Maxwell's
equations for the vector potential ${\bf A}(x,t)$ which yield
\begin{equation}
\partial_{tt}{\bf A}-c^2\partial_{xx}{\bf A}=
4\pi c \delta(x) {\bf J}({\bf A})+\dot \delta(t) {\bf A}(x,0)+\delta(t) \dot{\bf A}(x,0),
\label{AppI:Max}
\end{equation}
where $\delta (x)$ is the Dirac delta function and a dot denotes differentiation
with respect to time. The first term on the right
hand side of Eq. (\ref{AppI:Max}) describes the electric current in the foil
and the delta function, $\delta (x)$, represents its sharp localization.
The electric current ${\bf J}({\bf A})$ is a function of the vector
potential ${\bf A}(0,t)$ at $x=0$. The last two terms on the right hand
side of Eq. (\ref{AppI:Max}) are equivalent to the initial conditions:
${\bf A}(x,0)={\bf A}_0(x)$ and $\partial_{t}{\bf A}(x,0)=\dot {\bf A}_0(x)$.
Here the functions ${\bf A}_0(x)$ and $\dot {\bf A}_0(x)$ define the incident
electromagnetic wave, ${\bf A}_0(x,t)$. Convolution of the Green
function for the one-dimensional wave equation, $G(x,t;s,\tau)=\theta((t-\tau)-|x-s|/c)/2$, 
with the terms in the r.h.s. of Eq. (\ref{AppI:Max})
yields
\begin{equation}
{\bf A}(x,t)={\bf A}_0(x,t)+2 \pi \int_0^{t-|x|/c}{\bf J}({\bf A}(0,\tau))d\tau.
\label{AppI:A0xt}
\end{equation}
Assuming $x=0$ on
both sides of Eq. (\ref{AppI:A0xt}) and taking the derivative with respect to time,
we obtain
\begin{equation}
\dot{\bf A}(0,t)=\dot{\bf A}_0(0,t)+2 \pi {\bf J}({\bf A}(0,t)).
\label{AppI:A0t}
\end{equation}
On the right hand side, $\dot{\bf A}_0(0,t)$, is a known function 
and the electric current, ${\bf J}$, is assumed
to be a given function of ${\bf A}(0,t)$. In this way a nonlinear
boundary problem for a system of partial differential
equations is reduced to the ordinary differential equation  for
the field inside the foil (\ref{AppI:A0t}). Solving this equation we find the
vector potential ${\bf A}(0,t)$ inside the foil. Substituting it into Eq.
(\ref{AppI:A0xt}) we obtain the expression that describes transmitted
and reflected waves.

Taking into account the generalized electron momentum conservation, ${\bf p}-e{\bf A}/c=\,$constant,   
and the relationship between the electric current and the electron velocity, ${\bf J}=-e n_e l {\bf v}
=-e n_e l c {\bf p}/\sqrt{m_e c^2+p^2}$, where $n_e$ and $l$ are the electron density and the foil thickness,
we find that
the 1D equation (\ref{AppI:A0t}) for the stationary motion of a thin foil interacting
of its interaction with a rotating 
electric field can be written in the form

\begin{equation}
\tilde{u}_{2}=\varepsilon _0\tilde{u}_{3}/\gamma ,  \label{eq0App1} 
\end{equation}%
\begin{equation}
\tilde{u}_{3}=a-\varepsilon _0\tilde{u}_{2}/\gamma .  \label{eq0App2} 
\end{equation}
Here $\varepsilon _0$ is the dimensionless parameter \cite{32},
\begin{equation}
\varepsilon _0=\frac{2\pi e^2 n_el }{m_e \omega_0 c},  \label{eq0App3} 
\end{equation}
proportional to the surface electric charge of the foil: $e n_e l$.

Solving this system of algebraic equations we obtain
\begin{equation}
\tilde{u}_{2}=\frac{\varepsilon _0}{\sqrt{2}a}
\frac{ 
\sqrt{4 a^2+\left(1-a^2+\varepsilon _0^2\right)^2}-\left(1-a^2+\varepsilon _0^2\right)
}
{
\sqrt{4 a^2+\left(1-a^2+\varepsilon _0^2\right)^{2}}+\left(1+a^2-\varepsilon _0^2\right)
}
\label{eq0App4} 
\end{equation}%
\begin{equation}
\tilde{u}_{3}=\frac{1}{2 a}
\left[ 
\sqrt{4 a^2+\left(1-a^2+\varepsilon _0^2\right)^{2}}-\left(1-a^2+\varepsilon _0^2\right)
\right].  
\label{eq0App5} 
\end{equation}

In the limit of a relatively weak EM field when $\varepsilon_0 \gg a\gg 1 $ 
solutions to Eqs.(\ref{eq0App4}) 
and (\ref{eq0App5}) 
have the asymptotics
\begin{equation}
\tilde{u}_{2}=\frac{\varepsilon _0}{1+\varepsilon _0^2}a-
\frac{\varepsilon _0 (1-\varepsilon _0^2)}{2 (1+\varepsilon _0^2)^3}a^3+O(a^4),
\end{equation}%
\begin{equation}
\tilde{u}_{3}=\frac{1}{1+\varepsilon _0^2}a-\frac{\varepsilon _0^2}
{(1+\varepsilon _0^2)^3}a^3+O(a^4).
 \label{eq0App6} 
\end{equation}%

In the opposite limit for $a \gg \varepsilon_0\gg 1 $ the asymptotics are 
\begin{equation}
\tilde{u}_{2}=\varepsilon _0-\frac{\varepsilon _0 (1+\varepsilon _0^2)}{2 a^2}a+O(a^{-4}),
\end{equation}%
\begin{equation} 
\tilde{u}_{3}=a-\frac{\varepsilon _0^2}{a}+\frac{\varepsilon _0^2}{a^3}+O(a^{-4}).
 \label{eq0App7} 
\end{equation}%

As we can see, in the dissipative range of parameters, which corresponds to a relatively low EM field 
amplitude $\varepsilon_0 \gg a$, 
the component of the electron momentum, $p_{||}$, parallel to the electric field 
 is much larger than the perpendicular component, $p_{\perp}$. In the limit
of a strong electric field, $\varepsilon_0 \ll a$, we have $p_{||} \ll p_{\perp}$, 
i.e. the electron momentum 
is almost perpendicular to the
instantaneous direction of the electric field, contrary to the case of a  3D point electric charge, 
when the dissipative regime with 
$p_{||} \gg p_{\perp}$ requires the condition: $a \gg \varepsilon_{rad}$.

\section{DEPENDENCE OF RADIATION FRICTION EFFECTS ON THE PARAMETER ${\large \chi}$ 
IN THE ULTRARELATIVISTIC ELECTRON INTERACTION WITH AN EM PULSE}

Incorporating the radiation friction effects in the Landau-Lifshitz form into the electron equation 
of motion (\ref{eq01a}, 6),
\begin{equation}
\frac{d {\bf p}}{dt}=-e {\bf E}-\frac{e}{m_e c \gamma} \left({\bf p} \times {\bf B} \right)-
\frac{2 e^4}{3 m_e^4 c^7}\frac{{\bf p}}{\gamma}
\left[
\left( m_e c \gamma {\bf E}+{\bf p} \times {\bf B} \right)^2 -\left({\bf p}\cdot {\bf E}\right)^2
\right].
 \label{eq0App8} 
\end{equation}%
Here we retained the main order terms in the radiation friction force.
If $\varepsilon_{rad} a \gamma^2\gg 1$, which for a ten petawatt laser 
with $a=300$ corresponds to $\gamma\approx 500$, 
the interaction becomes purely dissipative and Eq. (\ref{eq0App8}) can be reduced to
\begin{equation}
\frac{d  p_1}{dt}=-\varepsilon_{rad} \omega a^2(-2t) \frac{p_1^2}{m_e c}.
\label{eq0App9a} 
\end{equation}%
In this equation we assume the head-collision case 
of the electron interaction with the laser pulse, for which
$x\approx -ct$ and $a(x-ct)\approx a(-2t)$. Its solution is given by 
\begin{equation}
p_1(t)=-\frac{p_{1,0} m_e c}{m_e c+\varepsilon_{rad} \omega p_{1,0}\displaystyle{\int_0^t a^2(-2t')dt'}}.
\label{eq0App9} 
\end{equation}%
For a pulse envelope, $a(t)$, of the Gaussian form, $a(t)=a_0 \exp \left(-t^2/2\tau_{las}^2\right)$,
Eq. (\ref{eq0App9}) takes the form
\begin{equation}
p_1(t)=-\frac{p_{1,0} m_e c}{m_e c+\varepsilon_{rad} a_0^2\omega \tau_{las} p_{1,0} \sqrt{\pi/32}
\left[\displaystyle{{\rm erf}(\sqrt{2}t/\tau_{las})-1}\right]},
\label{eq0App10} 
\end{equation}%
where ${\rm erf}(x)$ is the error function equal to \cite{37}
\begin{equation}
{\rm erf}(x)=\frac{\sqrt{\pi}}{2}\int_0^x \exp(-t^2)dt.
\end{equation}%
The dependence given by Eq. (\ref{eq0App10}) shows that for large enough $\omega \tau_{las} p_{1,0} a_0^2$ 
the electron momentum
tends to the limit of 
\begin{equation}
p_1(t)\to 
\sqrt{ 
\frac{\pi}{8}
} 
\frac{m_e c}
{\varepsilon_{rad} \omega \tau_{las} a_0^2}, 
\label{eq0App11} 
\end{equation}%
which is independent of initial momentum value, $p_{1,0}$.
In this limit, the parameter $\chi$ is equal to 
\begin{equation}
\chi= 
\sqrt{
\frac{\pi}{8}
}
\frac{1}
{\omega \tau_{las} a_0}. 
\label{eq0App12} 
\end{equation}%
Although for $\omega \tau_{las}\gg 1$ and $a_0\gg 1$ it is substantially smaller than unity, 
the QED parameter, $\chi_e$, being a factor 
$\alpha^{-1}\approx 137$ larger, can be larger than unity as shown in Ref. \cite{15nim}.

\end{document}